\documentclass[a4paper,11pt]{article}
\usepackage{amssymb,amsmath,amsfonts}

\newtheorem{theorem}{Theorem}

\newtheorem{definition}[theorem]{Definition}

\newtheorem{proposition}[theorem]{Proposition}

\begin{document}

\title{Temparature as Order Parameter \\
of Broken Scale Invariance}
\author{Izumi OJIMA\thanks{%
Research Institute for Mathematical Sciences, Kyoto University, Kyoto
606-8502, Japan}}
\date{}
\maketitle

\begin{abstract}
In algebraic quantum field theory the (inverse) temperature is shown to be a
macroscopic \textit{order parameter} to parametrize mutually disjoint thermal 
\textit{sectors} arising from the \textit{broken scale invariance} under 
renormalization-group transformations. 

This is accomplished in a mathematical formalism for the consistent
treatment of \textit{explicitly broken symmetries} such as broken scale
invariance, on the basis of a clear-cut criterion for the symmetry breakdown
in a unified scheme for sectors proposed recently by the author.
\end{abstract}

\section{Introduction}

The purpose of the present paper is to give the explanation and the proof of
the following statement:

\begin{theorem}
In the standard setting up of algebraic quantum field theory, the inverse
temperature $\beta :=(\beta ^{\mu }\beta _{\mu })^{1/2}$ is a macroscopic 
\textbf{order parameter }for parametrizing mutually \textit{disjoint}
sectors in the thermal situation arising from the \textbf{broken scale
invariance} under the renormalization-group transformations, where $\beta
^{\mu }$ is an inverse temperature 4-vector of a relativistic KMS state $%
\omega _{\beta ^{\mu }}$ describing a thermal equilibrium in its rest frame.
\end{theorem}

\noindent This is obtained in my recent attempts to facilitate smooth
accesses to the mathematical methods of algebraic quantum field theory (QFT
for short) for their use in the actual problems in physics in such\ forms as

a) a mathematical formulation of \textit{explicitly broken symmetries} such
as the above mentioned broken scale invariance, and,

b) an attempt for the mathematical understanding of mutual relations among
quantum, thermal and geometrical aspects embodied in QFT in various physical
contexts involving such key notions as renormalization group and order
parameters, etc.

In the following the above two results crucial for the explanation of the
main theorem are shown to be naturally understood in a unified scheme
proposed in my recent projects for controlling the mutual relations between
the micro- and macroscopic aspects involved in quantum physics from the
viewpoint of \textit{superselection structures}. For this purpose, first let
me start with a brief survey of the essence of the results obtained in \cite%
{Unif03}, in combination with some standard basic materials and tools
available in the algebraic QFT necessary for our purpose.

The proposed scheme in \cite{Unif03} is meant for treating general sector
structures on the basis of selection criteria. It has been extracted from a
general formulation of non-equilirium local states \cite{BOR01} and adapted
to the discussions of the sector structures arising from internal symmetries
without and with their spontaneous breakdown. In the original
Doplicher-Roberts (DR) theory \cite{DR90}, the global gauge group $G$ of an
internal symmetry and the field algebra $\mathfrak{F}$ consisting of $G$%
-non-singlet quantum fields $\hat{\varphi}^{i}(x)$ are shown to be recovered
from the algebra $\mathfrak{A}=\mathfrak{F}^{G}$ of $G$-singlet observables
in combination with the data on states of physical relevance on $\mathfrak{A}
$ selected by the Doplicher-Haag-Roberts (DHR) criterion \cite{DHR} as
states with localizable charges. Previously, this theory was satisfactorily
formulated only in the cases with \textit{unbroken} symmetries, and its
general significance was not properly understood owing to its mathematical
sophistication. In \cite{Unif03}, such general essence extracted from this
theory has been extended not only to the \textit{broken symmetries} but also
to such \textit{thermal situations} as involving non-equilibrium local
states in QFT \cite{BOR01}; through these results a unified theoretical
scheme to control micro-macro transitions is seen to emerge, on the basis of
categorical adjunctions as \textit{matching conditions} between selected
states and reference states for comparison which are realized by `solving'
the selection criteria. By these attempts, it has turned out that the
proposed scheme allows one to treat various different physical theories
specialized to each specific domain in nature, in quite a homogeneous and
unified way, such as QFT in the vacuum situation, statistical mechanics of
quantum fields at finite temperatures and their non-equilibrium states with
non-trivial space-time dependent structures and behaviours in the following
forms:

\begin{enumerate}
\item DHR-DR superselection theory \cite{DHR, DR90} and its reformulation 
\cite{Unif03}, \ 

\item extension to the situations with a spontaneously broken internal
symmetry \cite{Unif03},

\item general formulation of non-equilibrium local states in QFT \cite%
{BOR01, Oji02, Oji03}.
\end{enumerate}

According to a suitable choice of a selection criterion, we select all the 
\textit{physical states of relevance} to each physical domain or aspects of
interest to be described theoretically, from the set of all generic states
on the net of quantum local observables or the algebra of quantum fields.
The meaningful choices of the selection criteria have turned out (see \cite%
{Unif03}) to involve some \textit{standard reference systems}, such as the
group dual $\hat{G}$ in the above case of DHR-DR superselection theory
controlling the mutual relations between the observable algebra $\mathfrak{A}%
=\mathfrak{F}^{G}$ with a specified family $\mathcal{T}$ of physical states
with localizable charges and the non-commutative dynamical system $\mathfrak{%
F}\curvearrowleft G$ with the group of internal symmetry $G=Gal(\mathfrak{F}/%
\mathfrak{A})$ given as a Galois group. In the case of non-equilibrium local
states, such a role is played by the space $B_{K}$ of thermodynamic
parameters $(\beta ,\mu )$ and that $M_{+}(B_{K})$ of probability measures $%
\rho $ on $B_{K}$ describing the fluctuations of $(\beta ,\mu )$. Through
the comparison of a generic unknown state $\omega $ with members of standard
states $\omega _{\rho }=\mathcal{C}^{\ast }(\rho )$ equipped with parameters 
$\rho $ belonging to the reference system, we can judge whether $\omega $
satisfies the criterion or not. If the selection criterion and the standard
reference system are suitably set up, we can solve an \textquotedblleft
inverse\textquotedblright\ of a map $\mathcal{C}^{\ast }$ (\textit{c}$%
\rightarrow $\textit{q channel}) from the reference system to the set of
generic quantum states as a kind of \textquotedblleft left
adjoint\textquotedblright\ (\textit{q}$\rightarrow $\textit{c channel}) in
the categorical adjunction which provides us with the interpretation of a
generic selected state $\omega $ in terms of the vocabulary of the standard
known object $\rho $ belonging to the reference system. Here what plays
crucial roles is the relation (Sec.3) between the \textit{superselection
sectors} arising from the classification of states and representations of
the algebra of quantum physical variables and the \textit{spectrum of the
centre} (of the \textit{universal }representation containing all the
selected relevant ones), the latter of which plays the roles of \textit{%
macroscopic order parameters }to distinguish among different sectors on the
basis of their mutual disjointness.

If this kind of machinery works appropriately, then it will allow us to
analyze in terms of the selection criteria the mutual relations among
different theories to describe different physical domains or aspects, on the
basis of which we can attain a framework to allow one specific form of a
theory adapted to a fixed scale region in the physical world to be freely
transferred to another one, according to the changes in length scales and
aspects relevant to the problems in question. At this point, however, we
note that all the above results are obtained in the essential use of the
basic notions and mathematical techniques developed on the notions of vacuum
states (characterized by the spectrum condition) and/or KMS states (based on
the KMS condition), which have been adopted as they stand without being
re-viewed from the novel viewpoint of the proposed scheme. Namely, one of
the most important \textit{family of KMS states} (including the vacuum
states as a special case of $\beta =\infty $ or $T=0$) as standard reference
systems has been left untouched without having acquired a physically natural
reformulation. It is clear that we cannot attain the aim of a unified scheme
in a self-consistent way unless this family is successfully incorporated in
the framework briefly explained above.

For this purpose, we start to approach to this question by examining the
geometric and algebraic meanings of the temperatures, asking ourselves a
question, \textquotedblleft Is a temperature \textit{a priori parameter or
physical quantity}?\textquotedblright . As will be examined in the next
section, the most useful hint can be found in the famous Takesaki theorem 
\cite{Tak70} showing the mutual disjointness $\omega _{\beta _{1}}\overset{%
\shortmid }{\circ }\omega _{\beta _{2}}$ between KMS states at different
temperatures, $\beta _{1}\neq \beta _{2}$, for a quantum C*-dynamical system
with \textit{type III} representations in its KMS states.

The general and physical meanings contained in the mathematical notions of
disjointness will be explained in Sec.3 in combination with the related
important and basic notions such as quasi-equivalence of representations,
folia and the role of the spectrum of the centre in classifying mutually
disjoint representations and states. On the basis of these preparations, the
criterion for symmetry breakdown is presented in Sec.4. To adapt the above
ingredients to the situation with the broken scale invariance in thermal
situations,\textit{\ }an augmented algebra as a composite system of a
genuine quantum system and a classical macroscopic system is constructed in
Sec.5 to accommodate \textit{spontaneously and/or explicitly broken
symmetries }unifying the viewpoint presented in Sec.4 and the notion of
scaling nets and algebras due to Buchholz and Verch \cite{BucVer}. In use of
the mappings to relate states on the original quantum system and the
augmented one, the claimed result on the role of (inverse) temperature as an
order parameter of broken scale invariance is proved in Sec.6. The final
section summarizes these consequences and briefly mentions the related
interesting problems in this context to be further investigated.

\section{Temperature: \textit{a priori parameter or physical quantity?}}

When an object in thermal equilibrium with respect to its rest frame is 
\textit{moving} relative to our frame, we observe in it such non-equilibrium
features as heat current just for the kinematical reason. This means that
the thermal equilibrium is meaningful only in reference to the frame in
which the object is \textit{at rest}, and it can be shown\textit{\ }\cite%
{Oji86} to imply the \textit{spontaneous breakdown of Lorentz boost symmetry}
in the context of (special-)relativistic QFT. In this situation, a Lorentz 
\textit{4-vector }$\beta ^{\mu }$\textit{\ of inverse-temperature} is shown
to be a key member of necessary parameters for specifying a thermal
equilibrium state: 
\begin{equation}
\beta ^{\mu }=\beta u^{\mu }\in \overline{V_{+}},\text{ \ \ \ }\beta
:=(\beta ^{\mu }\beta _{\mu })^{1/2}=(k_{B}T)^{-1},
\end{equation}%
where $V_{+}$ denotes the (open) forward lightcone in the Minkowski space
defined by $V_{+}:=\{(x^{\mu })\in \mathbb{R}^{4};x\cdot x=x^{0}x^{0}-\vec{x}%
\cdot \vec{x}>0,x^{0}>0\}$, and $k_{B}$ and $T$ are the Boltzmann constant
and a temperature, respectively. Such a thermal state $\omega _{(\beta ^{\mu
})}$ parametrized by $\beta ^{\mu }$ is shown \cite{BroBuc} to be
characterized by a \textit{relativistic KMS condition}, a relativistic
extension of the standard KMS condition \cite{BraRob}, and will be called a
relativisitc KMS state or simply a KMS state hereafter.

From the above explanation, the timelike unit vector $u^{\mu }:=$ $\beta
^{\mu }/\beta $ ($u^{\mu }u_{\mu }=1$) is seen to describe a relative
velocity for specifying the rest frame in which the relativisitc KMS state $%
\omega _{(\beta ^{\mu })}$ exhibits its genuine thermal equilibrium nature,
and it represents an \textit{order parameter} associated with this
spontaneous symmetry breaking (SSB) of Lorentz boosts \cite{Oji86}. In the
context of non-equilibrium local states \cite{BOR01}, a non-trivial
spacetime-dependent temperature distribution $x\longmapsto \beta ^{\mu
}(x)\in V_{+}$ is allowed to appear, in which $x\longmapsto u^{\mu }(x)=$ $%
\beta ^{\mu }(x)/\beta (x)$ becomes a time-like member of the vierbein field
to specify the rest \textit{frame} at each spacetime point $x$. Putting this
geometric aspect in a more general context of QFT formulated in a curved
spacetime (i.e., non-equilibrium states in a curved background spacetime 
\cite{Oji03, ISQM86}), we will encounter interesting mathematical-physical
problems at the boundary of geometry and thermodynamics.

The problem to be discussed here is, however, a question concerning another
factor $\beta =(\beta ^{\mu }\beta _{\mu })^{1/2}$ in the formula $\beta
^{\mu }=\beta u^{\mu }\in \overline{V_{+}}$, the inverse temperature itself
as a Lorentz scalar: is $\beta $ any kind of \textit{order parameter}
related with a certain symmetry similarly to the case of $u^{\mu }$? What is
suggestive in this context is the following famous theorem due to Takesaki 
\cite{Tak70} (see, for instance \cite{BraRob}):

\begin{theorem}[Takesaki]
Let $(\mathfrak{A}\underset{\alpha }{\curvearrowleft }\mathbb{R})$ be a
C*-dynamical system, and suppose that $\omega _{1}$ and $\omega _{2}$ are
KMS-states corresponding to two different values $\beta _{1}$, $\beta
_{2}\in \mathbb{R}.$ Assume that $\pi _{\omega _{1}}(\mathfrak{A})^{\prime
\prime }$ is a type-III von Neumann algebra. \newline
It follows that the states $\omega _{1}$ and $\omega _{2}$ are disjoint.
\end{theorem}

This mathematical fact suggests the following physical picture for such
quantum systems as QFT with infinite degrees of freedom, which intrinsically
involve type-III von Neumann algebras (as representation-independent local
subalgebras and/or in thermal situations): according to standard results 
\cite{DixC*, Sew} in the representation theory of C*-algebras, a family of 
\textit{disjoint} representations generate a \textit{non-trivial centre} in
the representation containing them as subrepresentations, whose elements can
be regarded as \textit{macroscopic order parameters }because of their mutual
commutativity and of their role in discriminating among different
representations within the family according to its spectrum. Since all the
GNS representations of the KMS states of QFT except for $\beta =0,\infty $
are known to be type-III von Neumann algebras, we are naturally led to a
situation with \textit{continuous superselection sectors} formed by KMS
states at different temperatures, distinguished mutually by \textit{%
macroscopic central observables} (in a representation containing all the KMS
states) among which the (inverse) temperature $\beta $ is found. Namely, $%
\beta $ becomes in this situation a \textit{physical macro-variable} running
over the space of all possible thermal equilibria, instead of being an a
priori given fixed parameter as is treated in the standard approach to
statistical mechanics.

Starting from this observation, it will be shown in the following that $%
\beta $ is a physical \textit{order parameter} corresponding to\textit{\ }%
the \textit{spontaneously} or \textit{explicitly} \textit{broken} \textit{%
scale invariance}\ under the renormalization-group transformations; namely, $%
\beta $'s not only parametrize continuous sectors of thermal equilibria at
different temperatures, but also are mutually interrelated by the
renormalization-group transformation associated with the broken scale
invariance, which clarifies the geometric structure of the thermodynamic
classifying space identified with the spectrum of the above centre of the
representation universal within the KMS family.

\section{Classification of representations and states: central spectrum as
classifying space of sectors}

Just for convenience' sake, let us briefly recall and summarize the basic
mathematical notions relevant to the present context in the following form.

\bigskip

\noindent1) Folium / disjointness / quasi-equivalence:

Let $\mathfrak{A}$ be a unital C*-algebra and denote $E_{\mathfrak{A}}$ the
set of all states on $\mathfrak{A}$ defined as normalized positive linear
functionals. All representations $(\pi ,\mathfrak{H})$ are to be understood
here as unital *-representations in the sense that $\pi (\mathbf{1})=\mathbf{%
1}_{\mathfrak{H}}$, $\pi (A^{\ast })=\pi (A)^{\ast }$.

\begin{definition}[Folium]
Given a representation $(\pi ,\mathfrak{H})$ of $\mathfrak{A}$, a state $%
\varphi \in E_{\mathfrak{A}}$ is called $\pi $\textbf{-normal }if there
exists a density operator in $\mathfrak{H}$ s.t. $\varphi (A)=\mathrm{Tr}%
(\sigma \ \pi (A))$ ($\forall A\in \mathfrak{A}$). Totality $\mathfrak{f}%
(\pi )$ of $\pi $-normal states is called a \textbf{folium} of $\pi $:%
\begin{equation}
\mathfrak{f}(\pi ):=\{\mathfrak{A}\ni A\longmapsto \mathrm{Tr}(\sigma \ \pi
(A))\text{; }\sigma \text{: density operator in }\mathfrak{H}\}\text{.}
\end{equation}%
A folium $\mathfrak{f}(\omega )$ of a state $\omega \in E_{\mathfrak{A}}$ is
defined by $\mathfrak{f}(\omega ):=\mathfrak{f}(\pi _{\omega })$ w.r.t.~the
GNS representation $\pi _{\omega }$ of $\omega $.
\end{definition}

From the definition, it is clear that the linear span of a folium gives the
predual $(\pi (\mathfrak{A})^{\prime \prime })_{\ast }$ of the von Neumann
algebra $\pi (\mathfrak{A})^{\prime \prime }$, consisting of $\sigma $%
-weakly continuous linear functionals on $\pi (\mathfrak{A})^{\prime \prime
} $: $Lin(\mathfrak{f}(\pi ))=(\pi (\mathfrak{A})^{\prime \prime })_{\ast }$%
, $Lin(\mathfrak{f}(\pi ))^{\ast }=\pi (\mathfrak{A})^{\prime \prime }$. In
terms of this notion, the definitions of disjointness and quasi-equivalence
of representations can be understood in clear-cut way, as follows.

\begin{definition}[Disjointness]
\cite{DixC*} Two representations $(\pi _{1},\mathfrak{H}_{1}),$ $(\pi _{2},%
\mathfrak{H}_{2})$ of $\mathfrak{A}$ are said to be \textbf{disjoint} and
written as $\pi _{1}\overset{\shortmid }{\circ }\pi _{2}$, if and only if
they have \textbf{no} pair $\rho _{1},\rho _{2}$ of \textbf{unitarily
equivalent} non-trivial subrepresentations $0\neq \rho _{1}\prec \pi
_{1},0\neq \rho _{2}\prec \pi _{2}$, Likewise, states $\omega _{1},\omega
_{2}\in E_{\mathfrak{A}}$ with disjoint GNS representations $\pi _{\omega
_{1}}\overset{\shortmid }{\circ }\pi _{\omega _{2}}$ are said to be \textbf{%
disjoint} and written as $\omega _{1}\overset{\shortmid }{\circ }\omega _{2}$%
.
\end{definition}

According to the standard results (see \cite{DixC*}), the defined
disjointness is rephrased into the following equivalent forms:%
\begin{eqnarray}
\pi _{1}\overset{\shortmid }{\circ }\pi _{2} &\Longleftrightarrow &\mathfrak{%
f}(\pi _{1})\cap \mathfrak{f}(\pi _{2})=\emptyset  \notag \\
&\Longleftrightarrow &\mathcal{H}^{\mathfrak{A}}(\pi _{2}\leftarrow \pi
_{1})=0  \notag \\
&\Longleftrightarrow &c(P(\pi _{1}))\perp c(P(\pi _{2})),
\end{eqnarray}%
where $\mathcal{H}^{\mathfrak{A}}(\pi _{2}\leftarrow \pi _{1})$ is the set
of intertwiners from $\pi _{1}$ to $\pi _{2}$ defined by 
\begin{equation}
\mathcal{H}^{\mathfrak{A}}(\pi _{2}\leftarrow \pi _{1}):=\{T\in B(\mathfrak{H%
}_{1},\mathfrak{H}_{2})\text{; }T\pi _{1}(A)=\pi _{2}(A)T\text{ }(\forall
A\in \mathfrak{A})\}
\end{equation}%
and $P(\pi _{i})$ $\in \pi (\mathfrak{A})^{\prime }=\mathcal{H}^{\mathfrak{A}%
}(\pi \leftarrow \pi )$ ($i=1,2$) are projections corresponding to $\pi _{i}$
regarded as subrepresentations of a common representation $\pi $~(e.g., $\pi
=\pi _{1}\oplus \pi _{2}$) and $c(P)$ denotes the central support $%
c(P):=\min \{F$: projection $\in \pi (\mathfrak{A})^{\prime \prime }\cap \pi
(\mathfrak{A})^{\prime }\mathfrak{;}F\geq P\}$ of a projection $P$ ($\in \pi
(\mathfrak{A})^{\prime }$ or $\pi (\mathfrak{A})^{\prime \prime }$). If $\pi
_{1},\pi _{2}$ are both irreducible, disjointness $\pi _{1}\overset{%
\shortmid }{\circ }\pi _{2}$ means simply their unitary inequivalence. If
both $\pi _{1},\pi _{2}$ can be uniquely decomposed into irreducible
components, $\pi _{1},\pi _{2}$ have no common irreducible components.

The `opposite' situation to the disjointness is given by the notion of 
\textit{quasi-equivalence} defined next on the basis of the following
proposition:

\begin{proposition}[\protect\cite{DixC*}]
The following conditions for representations $\pi _{1},\pi _{2}$ of $%
\mathfrak{A}$ are all equivalent:

\begin{enumerate}
\item[(i)] No non-trivial subrepresentation of $\pi _{1}$ is disjoint from $%
\pi _{2}$ and no non-trivial subrepresentation of $\pi _{2}$ is disjoint
from $\pi _{1}$[\textbf{negation of disjointness}];

\item[(ii)] $\exists \Phi :\pi _{1}(\mathfrak{A})^{\prime \prime
}\rightarrow \pi _{2}(\mathfrak{A})^{\prime \prime }$: \textbf{isomorphism
of von Neumann algebras} s.t. $\pi _{2}(A)=\Phi (\pi _{1}(A))$ for $\forall
A\in \mathfrak{A}$;

\item[(iii)] $\exists n_{1},n_{2}$: cardinals s.t. $n_{1}\pi _{1}\cong
n_{2}\pi _{2}$ where $n_{1}\pi _{1}$ and $n_{2}\pi _{2}$ are, respectively,
multiples of $\pi _{1}$ and of $\pi _{2}$ [i.e., \textbf{unitary equivalence
up to multiplicities}];

\item[(iv)] $\mathfrak{f}(\pi _{1})=\mathfrak{f}(\pi _{2})$ for the \textbf{%
folia} of $\pi _{1},\pi _{2}$. \newline
\noindent If $\pi _{1},\pi _{2}$ are subrepresentations of a representation $%
\pi $ with the corresponding projections $P(\pi _{1}),P(\pi _{2})\in \pi (%
\mathfrak{A})^{\prime }$, the above (i)-(iv) are also equivalent to the next
(v):

\item[(v)] $c(P(\pi _{1}))=c(P(\pi _{2}))$ for the \textbf{central supports}
of $P(\pi _{1}),P(\pi _{2})$.
\end{enumerate}
\end{proposition}

\begin{definition}[Quasi-equivalence]
Two representations $\pi _{1},\pi _{2}$ satisfying one (and hence, all) of
the conditions (i)-(v) are said to be \textbf{quasi-equivalent}, and written
as $\pi _{1}\approx \pi _{2}$. States $\omega _{1}$ and $\omega _{2}$ are
said to be quasi-equivalent if the corresponding GNS representations $\pi
_{\omega _{1}}$ and $\pi _{\omega _{2}}$ are quasi-equivalent.
\end{definition}

\bigskip

\noindent 2) Pure phase vs. mixed phase; superselection \textit{sectors}
and\ \textit{order parameter}

While KMS states $\omega _{\beta }$ describing thermal equilibria are all 
\textit{mixed states }(except for the case of vacuum with $\beta =\infty $),
their decompositions into pure states are highly non-unique for quantum
dynamical systems with infinte degrees of freedom because of their
type-threeness. Therefore, it is more legitimate to understand a KMS state
as an entity in itself without reference to pure states. For this purpose,
we need to know the \textit{minimal units} of KMS states in order for a
generic one to be decomposed canonically. Such a unit is known to be found
in a thermodynamic\textit{\ pure phase }$\omega \in E_{\mathfrak{A}}$
characterized by \textit{ergodicity, }or equivalently by \textit{%
factoriality }defined by the \textit{triviality of centre} $\mathfrak{Z}%
_{\omega }(\mathfrak{A})=\pi _{\omega }(\mathfrak{A})^{\prime \prime }\cap
\pi _{\omega }(\mathfrak{A})^{\prime }=\mathbb{C}\mathbf{I}_{\mathfrak{H}%
_{\omega }}$ as a condition equivalent to the extremality in the set $%
K_{\beta }$\ of all KMS states at $\beta $ on $\mathfrak{A}\underset{\alpha }%
{\curvearrowleft }\mathbb{R}$ \cite{BraRob}.

Along this line, we call \textit{pure phases} any factor states $\omega $ or
factor representations with trivial centre. If a given state $\omega $ is 
\textit{not} a pure phase, it is called a \textit{mixed phase} whose GNS
representation $\pi _{\omega }$ has a \textit{non-trivial centre} $\mathfrak{%
Z}_{\omega }(\mathfrak{A})$\textit{. }As a commutative algebra, the centre $%
\mathfrak{Z}_{\omega }(\mathfrak{A})$ admits a \textquotedblleft
simultaneous diagonalization\textquotedblright\ due to the well-known
Gel'fand theorem expressing it as a function algebra $L^{\infty }(Spec(%
\mathfrak{Z}_{\omega }(\mathfrak{A})))$ on the spectrum $Spec(\mathfrak{Z}%
_{\omega }(\mathfrak{A}))$ consisting of characters or maximal ideals.
Corresponding to this, $\omega $ and $\pi _{\omega }$ are canonically
decomposed (= \textit{central decomposition}) into factor states (= pure
phases) and factor subrepresentations (= sectors): 
\begin{eqnarray}
\omega (A) &=&\int_{Sp(\mathfrak{Z}_{\omega }(\mathfrak{A}))}\omega
_{\lambda }(A)d\mu (\lambda ), \\
\pi _{\omega }(A) &=&\int_{Sp(\mathfrak{Z}_{\omega }(\mathfrak{A}))}^{\oplus
}\pi _{\omega _{\lambda }}(A)d\mu (\lambda ).
\end{eqnarray}

\bigskip

According to the above proposition, any pair of factor states or factor
representations are either\textit{\ quasi-equivalent }or\textit{\ disjoint}
and if\textit{\ }pure phases $\omega _{1},\omega _{2}$ are disjoint, there
exists $\exists C\in \mathfrak{Z}_{\pi }(\mathfrak{A})$ s.t. $\omega
_{1}(C)\neq \omega _{2}(C)$ within a representation $\pi $ containing $\pi
_{\omega _{1}}\oplus \pi _{\omega _{2}}$. In this sense, each point of $Spec(%
\mathfrak{Z}_{\omega }(\mathfrak{A}))$ represents a realized value of
parameters to distinguish among different pure phases contained in a mixed
phase $\omega $, and hence, each central element $C\in \mathfrak{Z}_{\omega
}(\mathfrak{A})$ can be identified with a \textit{macroscopic} \textit{order
parameter}.

Therefore, a mixed phase represents just the situation of a \textit{%
superselection rule} consisting of \textit{sectors} each of which is
identified with a folium of \textit{pure phases} or its factor
representations labelled by a point in $Spec(\mathfrak{Z}_{\omega }(%
\mathfrak{A}))$.

\bigskip

\noindent 3) Scheme for classifying microscopic sectors by means of
macroscopic order parameters

The above observations naturally lead to a \textquotedblleft unified scheme
for generalized sectors based upon \textbf{selection criteria}%
\textquotedblright \textbf{\ }\cite{Oji03} which can be schematized as
follows: \newline

i) $\left[ (q:)%
\begin{array}{c}
\text{generic objects} \\ 
\text{\textit{to be selected}}%
\end{array}%
\right] $ $\underset{\underset{\uparrow }{\uparrow }}{\Longrightarrow }$ii) $%
\left[ 
\begin{array}{c}
\text{\textit{standard reference} system with} \\ 
\text{\textit{classifying space} of \textit{sectors}}%
\end{array}%
(:c)\right] $

\ \ \ \ \ \ \ \ \ \ \ \ \ \ \ \ \ \ \ iii) \textit{comparison }of i) with ii)

$\ \ \ \ \ \ \ \ \ \ \ \ \ \ \ \ \ \ \ \Uparrow \ \ \ \ \ \ \ \ \ \ \ \ \ \
\ \ \ \ \ \ \ \ \ \ \ \ \ \ \ \ \ \ \ \Downarrow $

iv) $\left[ 
\begin{array}{c}
\text{\textit{selection criterion}:} \\ 
\text{ii)}\underset{\text{\textit{c-q channel}}}{\Longrightarrow }\text{i)}%
\end{array}%
\right] $\ $\underset{\text{\textit{adjunction}}}{\overset{\text{categorical}%
}{\rightleftarrows }}$ $\left[ 
\begin{array}{c}
\text{\textit{interpretation} of i) in} \\ 
\text{ terms of ii)}\newline
\text{: \ i)}\underset{\text{\textit{q-c channel}}}{\Longrightarrow }\text{%
ii)}%
\end{array}%
\right] $\newline

\bigskip

\noindent The applicability of this scheme has been confirmed in \cite%
{Unif03} in three major situations, the \textit{DHR superselection rule} 
\cite{DHR, DR90} to explain the operational origin of internal symmetries,
its extension to SSB cases \cite{Unif03} and the formulation of
non-equilibrium local states in QFT \cite{BOR01, Oji02, Oji03}. In all three
cases, the item i) is given by the set of states on the net $\mathcal{O}%
\longmapsto \mathfrak{A}(\mathcal{O})$ of local observables or its global
algebra $\mathfrak{A}$ to be explained in the next section (or, a version of
it slightly restricted w.r.t.~the energy spectrum in the case of
non-equilibrium local states \cite{Oji03, Unif03}). The item ii) is chosen
in the discussion of non-equilibrium local states the set $K:=Conv(\cup
_{\beta \in V_{+}}K_{\beta })$ of all the convex combinations of
(relativisitc) KMS states $\in K_{\beta }$ at all possible inverse
temperatures $\beta \in V_{+}$. In contrast, the corresponding choices in
the DHR-DR theory and its extension to SSB are not known a priori, which
turn out through the analyses to be, respectively, the group dual $\hat{G}$
(or functions $l^{\infty }(\hat{G})$ on it) of the group $G$ of the arising
internal symmetry and the sections $l^{\infty }(\amalg _{gH\in G/H}g\hat{H}%
g^{-1})$ of a fibre bundle related with the group $H$ of unbroken remaining
symmetry and that $G$ of the spontaneously broken symmetry. The item iii) is
just to compare a given unknown generic state $\omega $ in i) with those
special states in i) sent from the standard reference system ii) by the
embedding map as a c-q channel ii)$\Longrightarrow $i). If $\omega $ is
judged to be identified with one of such embedded reference states, then the
`inverse' map as a q-c channel i)$\Longrightarrow $ii) provides the \textit{%
interpretation} of $\omega $ in terms of ii). In this way, the $G$-charge
contents of $\omega $ in the DHR-DR case is described in terms of the
(fluctuation probability over) unitary equivalence classes $\subset \hat{G}$%
, and, in the SSB case, in terms of the data of $gH\in G/H$ specifying the
position (within the family of degenerate vacua parametrized by $G/H$) of a
vacuum sector (e.g., the spatial direction of the magnetization in the
example of a Heisenberg ferromagnet) to which $\omega $ belongs, in
combination with the $H$-charge contents of $\omega $. In the case of
non-equilibrium local states, $\omega $ is characterized as such a state if
it shows the agreement for certain restricted class of pointlike quantum
thermal observables (defined by some asymptotic limits of elements in the
local net $\mathcal{O}\longmapsto \mathfrak{A}(\mathcal{O})$ converging to a
spacetime point $x\in \mathbb{R}^{4}$) with a standard state of such a form
as $\omega _{\rho }=\int_{B_{K}}d\rho (\beta ,\mu )\omega _{\beta ,\mu }$
with $\rho \in M_{+}(B_{K})$ describing the statistical fluctuations of
thermodynamic parameters $(\beta ,\mu )\in B_{K}$ over various thermodynamic
pure phases. In this case, $d\rho (\beta ,\mu )$ gives the thermal
interpretation of a generic non-equilibrium state $\omega $ at a point $x$.

Since all these are in use of some systematic techniques available for
vacuum states and/or KMS states based upon their mathematical formulation,
it is quite important to try the possibility to treat all the KMS states
(including $\beta =\infty $ as a vacuum) themselves just according to the
same spirit as above.

\section{Criterion for symmetry breakdown}

To show that the inverse temperature appears as an order parameter of broken
scale invariance, we need to give precise formulations of

\begin{itemize}
\item[a)] the scale transformations which may possibly be allowed to involve 
\textit{explicit breaking} effects such as the presence of non-vanishing
mass terms,
\end{itemize}

\noindent and of

\begin{itemize}
\item[b)] a \textit{criterion for symmetry breakdown},
\end{itemize}

\noindent and then to exhibit

\begin{itemize}
\item[c)] the role of the inverse temperature as an order parameter for this
broken scale invariance.
\end{itemize}

For this purpose, we start here with the criterion for a spontaneous
symmetry breaking (SSB), as a special case of b) applicable to the
symmetries described by a strongly continuous \textit{automorphic action} $%
\tau $ of a locally compact group $G$ on the C*-algebra $\mathfrak{F}$ of
quantum fields: $G\underset{\tau }{\curvearrowright }\mathfrak{F}$. To be
more precise, the algebraic formulation of QFT is usually \textit{not} based
upon a C*-algebra $\mathfrak{F}$ of quantum \textit{fields} which are in
general not directly observable owing to their non-trivial behaviours under
the group $G$ of an internal symmetry. Instead, the basic ingredient to play
the principal roles is a\textit{\ net }$\mathcal{O}\longmapsto \mathfrak{A}(%
\mathcal{O})$ \textit{of} \textit{local observables} with each local
subalgebra $\mathfrak{A}(\mathcal{O})$ defined as a W*-algebra; any
self-adjoint element $A=A^{\ast }\in \mathfrak{A}(\mathcal{O})$ is supposed
to correspond to a physical quantity, experimentally observable within a
spacetime region $\mathcal{O}$ chosen from a suitable family of spacetime
regions which constitute a directed set w.r.t.~the inclusion relation, the
typical choice being the family $\mathcal{K}$ $:=\{(b+V_{+})\cap
(c-V_{+});~a,b\in \mathbb{R}^{4}\}$ of double cones in the Minkowski
spacetime $\mathbb{R}^{4}$. The common properties to be naturally satisfied
by this net are taken as follows (see, for instance, \cite{Haag}):

i) \textquotedblleft Isotony\textquotedblright : for $\mathcal{O}_{1},%
\mathcal{O}_{2}\in \mathcal{K}$, the inclusion relation $\mathcal{O}%
_{1}\subset \mathcal{O}_{2}$ should imply $\mathfrak{A}(\mathcal{O}%
_{1})\subset \mathfrak{A}(\mathcal{O}_{2})$, on the basis of which the
global algebra $\mathfrak{A}=\overline{\underset{\mathcal{K}\ni \mathcal{O}%
\nearrow \mathbb{R}^{4}}{\cup }\mathfrak{A}(\mathcal{O})}^{\left\vert
\left\vert \cdot \right\vert \right\vert }$ of observables can be defined as
the C*-inductive limit of all the local subalgebras $\mathfrak{A}(\mathcal{O}%
)$, $\mathcal{O}\in \mathcal{K}$.

ii) Relativistic covariance: the family $\mathcal{K}$ can be regarded as a
category consisting of objects as double cones $\mathcal{O}\in \mathcal{K}$
and of arrows $\mathcal{O}\overset{(a,\Lambda )}{\rightarrow }(a,\Lambda )%
\mathcal{O}$ defined by the elements $(a,\Lambda )\in \mathcal{P}%
_{+}^{\uparrow }=\mathbb{R}^{4}\rtimes L_{+}^{\uparrow }$ of the Poincar\'{e}
group acting on the Minkowski spacetime $\mathbb{R}^{4}$ and $\mathcal{K}$,
respectively, by $[(a,\Lambda )x]^{\mu }=\Lambda _{\nu }^{\mu }x^{\nu
}+a^{\mu }$, $(a,\Lambda )\mathcal{O}=\Lambda \mathcal{O}+a$. Then, the
local net $\mathcal{O}\longmapsto \mathfrak{A}(\mathcal{O})$ should
constitute a \textit{functor} $\mathfrak{A}:\mathcal{K}\rightarrow Alg$ \cite%
{ISQM86, Fre02} from the category $\mathcal{K}$ to the category $Alg$ of
W*-algebras whose arrows are (normal) *-homomorphisms:%
\begin{eqnarray}
\mathcal{K}\ni \mathcal{O} &\rightarrow &\mathfrak{A}(\mathcal{O})\in Alg 
\notag \\
(a,\Lambda )\downarrow & \circlearrowleft & \downarrow \alpha _{(a,\Lambda
)}:=\mathfrak{A}(a,\Lambda )  \notag \\
\Lambda \mathcal{O}+a &\rightarrow &\mathfrak{A}(\Lambda \mathcal{O}+a)
\end{eqnarray}

iii) Local commutativity (as a mathematical formulation of Einstein
causality): for \textit{spacelike separated} regions $\mathcal{O}_{1},%
\mathcal{O}_{2}\in \mathcal{K}$ (i.e., $(x-y)\cdot (x-y)<0$ ($\forall x\in 
\mathcal{O}_{1}$,$\forall y\in \mathcal{O}_{2}$)) the local subalgebras $%
\mathfrak{A}(\mathcal{O}_{1})$ and $\mathfrak{A}(\mathcal{O}_{2})$ are
commutative in the sense of $AB=BA$ ($\forall A\in \mathfrak{A}(\mathcal{O}%
_{1})$,$\forall B\in \mathfrak{A}(\mathcal{O}_{2})$).

The basic standpoint of the algebraic QFT (though not completely implemented
yet) is to regard the algebra $\mathfrak{F}$ of unobservable quantum fields
acted upon by a group $G$ of internal symmetry as a kind of mathematical
device constructed by the method of Galois extension from $\mathfrak{A}$ by
`solving some equations' identified with a suitably chosen criterion to
select out a family of physically relevant states \cite{DHR, DR90}. This
strategy has definite merits in providing a clear picture for the mutual
relations between the two aspects involving spacetime(=external) and
internal symmetries, treating the former in such a form as the spacetime
dependence of the net $\mathcal{O}\longmapsto \mathfrak{A}(\mathcal{O})$ and
the latter in terms of an abstract non-commutative version of Tannaka-Krein
duality between an internal symmetry group $G$ and the representation
category $Rep_{G}$ realized in the superselection sectors of $\mathfrak{A}$,
respectively, and then combining two aspects in the form of the dynamical
system $\mathfrak{F}\curvearrowleft G$. In the next section, the former
aspect in relation with the scale changes is focused by means of the above
net $\mathcal{O}\longmapsto \mathfrak{A}(\mathcal{O})$ of local observables.
Concerning the problem of a \textit{symmetry breaking}, what is to be in
focus is the \textit{global} aspects in which the differences of symmetries
between internal and external become largely irrelevant.

With this understanding, we treat, as the algebra of the system under
consideration, the C*-algebra $\mathfrak{F}$ of qunatum fields acted upon by
a group $G$ which is supposed to represent all the possible kinds of
symmetries characterizing the physical system. In the case of \textit{%
unbroken} internal symmetry, the emerging group $G$ is verified to be a
compact Lie group, whereas there is no guarantee of such a characterization
of $G$ in the case of SSB. However, we assume here for technical reasons
such a restriction on $G$ that it should be a locally compact group.

Now, the traditional treatment of SSB is just based on the so-called
Goldstone commutators expressing the non-invariance $\omega (\delta (F))\neq
0$ ($\exists F\in \mathfrak{F}$) of a state $\omega $ under the
infinitesimal transformations $\delta $($\in Der(\mathfrak{F})$: densely
defined *-derivations on $\mathfrak{F}$) in the directions of symmetry
breaking; while this is a necessary condition for SSB in a pure (or more
generally, factor) state, its sufficiency can be assured only for spatially
homogeneous states such as vacuum ones. For instance, if factoriality of the
representation is not required, one can easily obtain a $G$-invariant state
even in the situation of SSB by averaging over a $G$-noninvariant factor
state, which evades the necessity of $\omega (\delta (F))\neq 0$. On the
other hand, such a representation $(\pi ,\mathfrak{H})$ can exist that in
spite of the absence of $G$-invariant states in $\mathfrak{H}$ the symmetry
is descibed by a unitary representation $G\ni g\longmapsto U(g)\in \mathcal{U%
}(\mathfrak{H})$ satisfying the so-called covariance condition $\pi (\tau
_{g}(F))=U(g)\pi (F)U(g)^{\ast }$. A general criterion for SSB can be given
in the following form of definition so as to avoid these kinds of
complications and to incorporate more general situations in an intrinsic way:

\begin{definition}
\cite{Oji03} A symmetry described by a strongly continous automorphic action 
$\tau $ of a locally compact group $G$ on a (global) C*-algebra $\mathfrak{F}
$ of quantum fields is said to be \textbf{unbroken }in a given
representation $(\pi ,\mathfrak{H})$ of $\mathfrak{F}$ if the spectrum of
centre $\mathfrak{Z}_{\pi }(\mathfrak{F}):=\pi (\mathfrak{F)}^{\prime \prime
}\cap $ $\pi (\mathfrak{F)}^{\prime }$ is pointwise invariant ($\mu $-a.e.
w.r.t.~the central measure $\mu $ for the central decomposition of $\pi $
into factor representations) under the action of $G$ induced on $Spec(%
\mathfrak{Z}_{\pi }(\mathfrak{F}))$. If the symmetry is not unbroken in $%
(\pi ,\mathfrak{H})$, it is said to be \textbf{broken spontaneously} there.
\end{definition}

This definition exhibits the essence of SSB\ as the \textit{conflict between
factoriality of }$(\pi ,\mathfrak{H})$ \textit{and unitary implementability
of }$G$ in it; in the usual approaches, the former point is respected at the
expense of the latter. Taking the opposite choice to respect
implementability we are led to the \textbf{non-trivial centre} which
provides convenient tools for analyzing \textbf{sector structure} and
flexible treatment of macroscopic \textbf{order parameters} to distinguish
different sectors, as explained in the previous section: the central
spectrum $Spec(\mathfrak{Z}_{\pi }(\mathfrak{F}))$ physically means
macroscopic order parameters\textit{\ }appearing in \textit{low-energy
infrared modes}, and hence, the physical essence of this definition can be
found in the\textquotedblleft \textit{infrared instability\textquotedblright 
} of the representation $(\pi ,\mathfrak{H})$\textit{\textbf{\ }}along the
direction of $G$-action which is in harmony with the intuitive physical
picture of SSB.

Note, however, that this definition admits the coexistence of unbroken and
broken \textit{sub}representations in a given representation $\pi $
suffering from SSB, according to which the central spectrum $Spec(\mathfrak{Z%
}_{\pi }(\mathfrak{F}))$ can further be decomposed into $G$-invariant
subdomains. In view of the requirement of $G$-invariance, each such \textit{%
minimal} subdomain is characterized by $G$-\textit{ergodicity,} which means 
\textit{central ergodicity} in the whole system. Therefore, $\pi $ is
decomposed into the direct sum (or, direct integral) of \textit{unbroken
factor representations} and \textit{broken non-factor representations}, each
component of which is centrally $G$-ergodic. In this way we obtain a \textit{%
phase diagram\ }on the spectrum of centre.

We can now construct a \textit{covariant} representation of $(\mathfrak{F}%
\underset{\tau }{\curvearrowleft }G)$ implementing broken $G$ \textit{%
minimally} in the sense of \textit{central }$G$-\textit{ergodicity} as
follows:\bigskip

\noindent1) Let $H$ be the \textit{maximal} closed subgroup of $G$ \textit{%
unbroken} in $(\pi ,\mathfrak{H)}$ with a covariant representation $(\pi ,U,%
\mathfrak{H)}$ of a C*-dynamical system $\mathfrak{F}\underset{\tau
\upharpoonright _{H}}{\curvearrowleft }H$ satisfying $\pi (\tau
_{h}(F))=U(h)\pi (F)U(h)^{\ast }$ for $\forall h\in H$. An augmented algebra 
$\mathfrak{\hat{F}}:=\mathfrak{F}\rtimes (H\backslash G)$ \cite{Unif03} is
defined by a C*-crossed product of $\mathfrak{F}$ with the homogeneous space 
$H\backslash G$ which is realized as the algebra of continuous cross
sections of C*-algebra bundle $G\times _{H}\mathfrak{F}\rightarrow
H\backslash G$:%
\begin{equation}
\mathfrak{\hat{F}}=\mathfrak{F}\rtimes (H\backslash G)=\Gamma (G\times _{H}%
\mathfrak{F}).
\end{equation}%
This can conveniently be identified with the algebra $C_{H}(G,\mathfrak{F})$
of $H$-equivariant continuous functions $\hat{F}$ on $G$ satisfying the
condition 
\begin{equation}
\hat{F}(hg)=\tau _{h}(\hat{F}(g))\text{ \ \ \ for }\forall g\in G,\forall
h\in H.
\end{equation}%
In what follows the identification of a cross section of $G\times _{H}%
\mathfrak{F}\rightarrow H\backslash G$ with an $H$-equivariant continuous
function on $G$ is always understood and, without changing the notation, we
freely move from one version to another. The product structure of $\mathfrak{%
\hat{F}}$ is simply given by the pointwise product, 
\begin{equation}
(\hat{F}_{1}\hat{F}_{2})(\dot{g}):=\hat{F}_{1}(\dot{g})\hat{F}_{2}(\dot{g}),
\end{equation}%
for $\hat{F}_{1},\hat{F}_{2}\in \mathfrak{\hat{F}}$, $\dot{g}\in H\backslash
G$, which is equivalent to $(\hat{F}_{1}\hat{F}_{2})(g):=\hat{F}_{1}(g)\hat{F%
}_{2}(g)$ in the version of $H$-equivariant continuous functions on $G$
consistently with the constraint of $H$-equivariance: $(\hat{F}_{1}\hat{F}%
_{2})(hg)=\hat{F}_{1}(hg)\hat{F}_{2}(hg)=\tau _{h}(\hat{F}_{1}(g))\tau _{h}(%
\hat{F}_{2}(g))=\tau _{h}((\hat{F}_{1}\hat{F}_{2})(g))$. The action $\hat{%
\tau}$ of $G$ on $\hat{F}\in \mathfrak{\hat{F}}$ is defined by 
\begin{equation}
\lbrack \hat{\tau}_{g}(\hat{F})](\dot{g}_{1})=\hat{F}(\dot{g}_{1}g),
\end{equation}%
or equivalently, $[\hat{\tau}_{g}(\hat{F})](g_{1})=\hat{F}(g_{1}g)$ for $H$%
-equivariant functions. The fixed-point subalgebra $\mathfrak{\hat{F}}^{G}$
of $\mathfrak{\hat{F}}$ under the action $\hat{\tau}$ of $G$ is given by the
constant section $\hat{F}:g\longmapsto F\in \mathfrak{F}^{H}$ because of the 
$H$-equvariance condition: $F=\hat{F}(hg)=\tau _{h}(\hat{F}(g))=$ $\tau
_{h}(F)$: 
\begin{equation}
\mathfrak{\hat{F}}^{G}\cong \mathfrak{F}^{H}.
\end{equation}

Then from a representation $(\pi ,U,\mathfrak{H)}$ of a C*-dynamical system $%
\mathfrak{F}\underset{\tau \upharpoonright _{H}}{\curvearrowleft }H$, a
representation $(\hat{\pi},\mathfrak{\hat{H}})$ is induced of the crossed
product $\mathfrak{\hat{F}}$ in the following way.\bigskip

\noindent2) With the left-invariant Haar measure $d\xi $ on $G/H$ (with 
\textit{left }$G$-action), the Hilbert space $\mathfrak{\hat{H}}$ is given
by $L^{2}$-sections of $G\times _{H}\mathfrak{H}$:%
\begin{equation}
\mathfrak{\hat{H}}=\int_{\xi \in G/H}^{\oplus }(d\xi )^{1/2}\mathfrak{H}%
=\Gamma _{L^{2}}(G\times _{H}\mathfrak{H},d\xi ),
\end{equation}%
which can be identified with the $L^{2}$-space of $\mathfrak{H}$-valued $%
(U,H)$-equivariant functions $\psi $ on $G$, 
\begin{equation}
\psi (gh)=U(h^{-1})\psi (g)\text{ \ \ \ for }\psi \in \mathfrak{\hat{H}}%
\text{, }g\in G\text{, }h\in H.
\end{equation}%
On this $\mathfrak{\hat{H}}$, the representations $\hat{\pi}$ and $\hat{U}$
of $\mathfrak{\hat{F}}$ and $G$ are defined, respectively, by 
\begin{align}
(\hat{\pi}(\hat{F})\psi )(g)& :=\pi (\hat{F}(g^{-1}))(\psi (g))\text{\ for }%
\hat{F}\in \mathfrak{\hat{F},}\psi \in \mathfrak{\hat{H},}g\in G, \\
(\hat{U}(g_{1})\psi )(g)& :=\psi (g_{1}^{-1}g)\text{ \ \ \ for }g,g_{1}\in G,
\end{align}%
and satisfy the covariance relation: 
\begin{equation}
\hat{\pi}(\hat{\tau}_{g}(\hat{F}))=\hat{U}(g)\hat{\pi}(\hat{F})\hat{U}%
(g)^{-1}.
\end{equation}%
\bigskip

\noindent3) $\mathfrak{F}$ is embedded into $\mathfrak{\hat{F}}$ by $\hat{%
\imath}_{H\backslash G}:\mathfrak{F}\hookrightarrow \mathfrak{\hat{F}}$
given by 
\begin{equation}
\lbrack \hat{\imath}_{H\backslash G}(F)](g):=\tau _{g}(F),
\end{equation}%
which is consistent with the $H$-equivariance condition: $[\hat{\imath}%
_{H\backslash G}(F)](hg)=\tau _{hg}(F)=\tau _{h}([\hat{\imath}_{H\backslash
G}(F)](g))$. This embedding map intertwines the $G$-actions $\tau $ on $%
\mathfrak{F}$ and $\hat{\tau}$ on $\mathfrak{\hat{F}}$, 
\begin{equation}
\hat{\imath}_{H\backslash G}\circ \tau _{g}=\hat{\tau}_{g}\circ \hat{\imath}%
_{H\backslash G}\text{\ \ \ }(\forall g\in G),
\end{equation}%
and hence, we have 
\begin{equation}
\lbrack \hat{\imath}_{H\backslash G}(\mathfrak{F})]^{G}=\hat{\imath}%
_{H\backslash G}(\mathfrak{F}^{G})\subset \hat{\imath}_{H\backslash G}(%
\mathfrak{F}^{H})=\mathfrak{\hat{F}}^{G}.
\end{equation}%
The mutual relations among (sub)algebras and mappings can be depicted by

\begin{equation}
\begin{array}{ccc}
& \mathfrak{\hat{F}}=\Gamma (G\underset{H}{\times }\mathfrak{F}) &  \\ 
^{\hat{\imath}_{G}}\nearrow \swarrow _{\hat{m}_{G}} &  & _{\hat{\imath}%
_{H\backslash G}}\nwarrow \searrow _{\hat{m}_{H\backslash G}} \\ 
\mathfrak{F}^{H}\cong \hat{\imath}_{H\backslash G}(\mathfrak{F}^{H})=%
\mathfrak{\hat{F}}^{G} & \overset{i_{H}}{\underset{m_{H}}{\rightleftarrows }}
& \mathfrak{F} \\ 
_{i_{G/H}}\nwarrow \searrow _{m_{G/H}} &  & ^{i_{G}}\nearrow \swarrow
_{m_{G}} \\ 
& \mathfrak{F}^{G} & 
\end{array}%
,
\end{equation}%
where the maps $i_{G}$ and $m_{G}$, etc., are, respectively, the embedding
maps (of a C*-algebra into another) and the conditional expectations defined
as operator-valued weights to extract fixed points, such as 
\begin{equation}
m_{G/H}:\mathfrak{F}^{H}\ni B\longmapsto m_{G/H}(B):=\int_{G/H}d\dot{g}\
\tau _{g}(B)\in \mathfrak{F}^{G}\mathfrak{.}
\end{equation}

\bigskip

\noindent4) Combining $\hat{\imath}_{H\backslash G}$ with $\hat{\pi}$, we
obtain a covariant representation $(\bar{\pi},\hat{U},\mathfrak{\hat{H}})$, $%
\bar{\pi}:=\hat{\pi}\circ \hat{\imath}_{H\backslash G}$, of $\mathfrak{F}%
\underset{\tau }{\curvearrowleft }G$ defined on $\mathfrak{\hat{H}}$ by 
\begin{equation}
(\bar{\pi}(F)\psi )(g):=\pi (\tau _{g^{-1}}(F))\psi (g)\text{ \ \ \ \ }(F\in 
\mathfrak{F},\psi \in \mathfrak{\hat{H}})
\end{equation}%
and satisfying 
\begin{equation}
\bar{\pi}(\tau _{g}(F))=\hat{U}(g)\bar{\pi}(F)\hat{U}(g)^{-1}.
\end{equation}

\bigskip

\noindent5) The sector structure is determined by the following information
on the relevant centres:

\begin{proposition}
When the von Neumann algebra $\pi (\mathfrak{F})^{\prime \prime }$ has a
trivial centre,$\ \mathfrak{Z}_{\pi }(\mathfrak{F})=\mathbb{C}\mathbf{1}$,
the centres of $\hat{\pi}(\mathfrak{\hat{F}})^{\prime \prime }$ and $\bar{\pi%
}(\mathfrak{F})^{\prime \prime }$ are given by 
\begin{equation}
\mathfrak{Z}_{\bar{\pi}}(\mathfrak{F})=L^{\infty }(H\backslash G;d\dot{g})=%
\mathfrak{Z}_{\hat{\pi}}(\mathfrak{\hat{F}}).
\end{equation}
\end{proposition}

This can be seen as follows: From the definition of $\mathfrak{\hat{F}}%
\subset C(H\backslash G,\mathfrak{F})\cong C(H\backslash G)\bar{\otimes}%
\mathfrak{F}$ it is clear that the commutative algebra $C(H\backslash G)$ is
contained in the centre of the C*-algebra $\mathfrak{\hat{F}}$. If this
centre is bigger than $C(H\backslash G)$, it contains a function $\hat{F}$
on $H\backslash G$ whose image $\hat{F}(\dot{g})$ at some point $\dot{g}\in
H\backslash G$ is not be a scalar multiple of the identity, which does not
commute with some element $F_{1}\in \mathfrak{F}$ because of the triviality
of the centre of $\mathfrak{F}$ due to $\mathfrak{Z}_{\pi }(\mathfrak{F})=%
\mathbb{C}\mathbf{1}$: $[\hat{F}(\dot{g}),F_{1}]\neq 0$.

In view of 3), $F_{1}$ can be embedded in $\mathfrak{\hat{F}}$ satisfying $%
\hat{\imath}_{H\backslash G}(F_{1})(\dot{e})=F_{1}$, and hence, we have $%
\hat{\tau}_{g^{-1}}\hat{\imath}_{H\backslash G}(F_{1})(\dot{g})=F_{1}$,
which shows the relation $[\hat{F},\hat{\tau}_{g^{-1}}\hat{\imath}%
_{H\backslash G}(F_{1})](\dot{g})=[\hat{F}(\dot{g}),F_{1}]\neq 0$. Thus, we
have $\mathfrak{Z}(\mathfrak{\hat{F}})=C(H\backslash G)$. Using the similar
arguments for $\hat{\pi}(\mathfrak{\hat{F}})^{\prime \prime }$ combined with 
$\mathfrak{Z}_{\bar{\pi}}(\mathfrak{F})\subset L^{\infty }(H\backslash G,d%
\dot{g})\bar{\otimes}\pi (\mathfrak{F})^{\prime \prime }$, we see $\mathfrak{%
Z}_{\hat{\pi}}(\mathfrak{\hat{F}})=L^{\infty }(H\backslash G;d\dot{g})$. The
equality $\mathfrak{Z}_{\bar{\pi}}(\mathfrak{F})=L^{\infty }(H\backslash G;d%
\dot{g})$ comes from the mutual disjointness $\pi \overset{\shortmid }{\circ 
}(\pi \circ \tau _{g})$ for $g\in G\diagdown H$ and $\mathfrak{Z}_{\pi }(%
\mathfrak{F})=\mathbb{C}\mathbf{1}$.

Since the homogeneous space $H\backslash G$ as the spectrum of the centre $%
\mathfrak{Z}_{\bar{\pi}}(\mathfrak{F})$ is transitive under the right action 
$G$ which is just the action induced on the central spectrum from $\hat{\tau}
$, the representation $(\bar{\pi},\mathfrak{\hat{H})}$ of the dynamical
system $\mathfrak{F}\underset{\tau }{\curvearrowleft }G$ is centrally $G$%
-ergodic, to which the criterion for SSB can be applied.

Adapting the above formulation to the GNS representation $(\pi =\pi _{\beta
},\mathfrak{H}=\mathfrak{H}_{\beta })$ of a KMS state $\omega _{\beta
=(\beta ^{\mu })}$ with $H=\mathbb{R}^{4}\rtimes SO(3)$, $G=\mathbb{R}%
^{4}\rtimes L_{+}^{\uparrow }$, we can reproduce the results on the SSB of
Lorentz boosts: $\mathfrak{Z}_{\bar{\pi}}(\mathfrak{F})=L^{\infty
}(SO(3)\backslash L_{+}^{\uparrow })=L^{\infty }(\mathbb{R}^{3})$ through
the identification $\beta ^{\mu }/\sqrt{\beta ^{2}}=u^{\mu }=(\frac{1}{\sqrt{%
1-\mathbf{v}^{2}/c^{2}}},\frac{\mathbf{v}}{\sqrt{1-\mathbf{v}^{2}/c^{2}}}%
)\longleftrightarrow \mathbf{v\in }\mathbb{R}^{3}$.

\section{How to formulate broken scale invariance?}

As noted at the beginning, the above discussion of symmetry breakdown was
concerning the \textit{spontaneous }breakdown of a symmetry described by a
group $G$ acting on the field algebra $\mathfrak{F}$ by \textit{automorphisms%
}. In contrast, the notion of the broken scale invariance is usually
understood in a physical system with such \textit{explicit breaking terms }%
as non-vanishing \textit{mass}, which seem to cause\textit{\ difficulties in
treating scale transformations as automorphisms} acting on the algebra
describing the system. However, the results on the \textit{scaling algebra}
in algebraic QFT due to Buchholz and Verch \cite{BucVer} shows that the
above negative anticipation can safely be avoided.

Their results can be summarized as follows. Let the following requirements
be imposed on all the possible renormalization-group transformations $%
R_{\lambda }$:

(i) $R_{\lambda }$ should map the given net $\mathcal{O}\rightarrow 
\mathfrak{A}(\mathcal{O})$ of local observables at spacetime scale $1$ onto
the corresponding net $\mathcal{O}\rightarrow \mathfrak{A}_{\lambda }(%
\mathcal{O}):=\mathfrak{A}({\lambda }\mathcal{O})$ at a scale $\lambda $,
i.e., 
\begin{equation}
R_{\lambda }:\,\mathfrak{A}(\mathcal{O})\rightarrow \mathfrak{A}_{\lambda }(%
\mathcal{O})
\end{equation}%
for every region $\mathcal{O}\subset \mathbb{R}^{4}$. Since both time and
space are scaled by the same $\lambda $, the light velocity $c$ remains
unchanged as their ratio.

(ii) In the Fourier-transformed picture, the subspace $\widetilde{\mathfrak{A%
}}(\widetilde{\mathcal{O}})$ of all (quasi-local) observables carrying
energy-momentum in the set $\widetilde{\mathcal{O}}\subset \mathbb{R}^{4}$
is transformed as 
\begin{equation}
R_{\lambda }:\,\widetilde{\mathfrak{A}}(\widetilde{\mathcal{O}})\rightarrow 
\widetilde{\mathfrak{A}}_{\lambda }(\widetilde{\mathcal{O}}),
\end{equation}%
where $\widetilde{\mathfrak{A}}_{\lambda }(\widetilde{\mathcal{O}}):=%
\widetilde{\mathfrak{A}}(\lambda ^{-1}\widetilde{\mathcal{O}})$. In view of
the duality between spacetime coordinates $x^{\mu }$ and energy-momenum $%
p_{\mu }$ involved in the Fourier transformation, this requirement implies
the invariance of the quantity $p_{\mu }x^{\mu }=Et-\vec{p}\cdot \vec{x}$
called \textquotedblleft action\textquotedblright\ in physics, as a
consequence of which the Planck constant $\hbar $ with the dimension of
action also remains invariant.

(iii) $R_{\lambda }$ should be bounded continuous maps uniformly in $\lambda 
$ (even if they may not be isomorphisms): concerning the possibility of
non-isomorphisms, Buchholz and Verch in \cite{BucVer} remark
\textquotedblleft In the case of dilation invariant theories the
transformations $R_{\lambda }$ are expected to be isomorphisms, yet this
will not be true in general since the algebraic relations between
observables may depend on the scale.\textquotedblright\ In contrast to their
focus on the high-energy limits in the context of vacuum situations, our
interest here is in the thermal situations involving all the possible
temperatures. But the similar point to the scale dependence of the basic
algebraic relations should be expected to show up especially in the
direction to the low temperature side, because of the increasing complexity
of phase structures arising from the bifurcating processes of phase
transitions. In view of the seemingly ad hoc choices of the starting
dynamics and the algebras of relevant physical variables in the standard
approaches to phase transitions, it should be certainly one of the
non-trivial important questions \textit{whether all the variety of different
thermodynamic phases can be totally attributed to that of the realized
states and representations of one and the same fixed dynamical system} with
a fixed algebra of observables and a fixed dynamics acting on the former.

Then, the \textit{scaling net }$\mathcal{O}\rightarrow \mathfrak{\hat{A}}(%
\mathcal{O})$ corresponding to the original local net $\mathcal{O}%
\rightarrow \mathfrak{A}(\mathcal{O})$ of observables is defined as the
local net consisting of scale-changed observables under the action of all
the possible choice of $R_{\lambda }$ satisfying (i)-(iii). With the
derivation process referred to \cite{BucVer}, the obtained results on the
structure of $\mathfrak{\hat{A}}(\mathcal{O})$ can be reformulated into the
identification of $\mathfrak{\hat{A}}(\mathcal{O})$ with the algebra $\Gamma
(\amalg _{\lambda \in \mathbb{R}^{+}}\mathfrak{A}_{\lambda }(\mathcal{O}))$
of sections $\mathbb{R}^{+}\ni \lambda \longmapsto \hat{A}(\lambda )\in 
\mathfrak{A}_{\lambda }(\mathcal{O})$ of algebra bundle $\amalg _{\lambda
\in \mathbb{R}^{+}}\mathfrak{A}_{\lambda }(\mathcal{O})\twoheadrightarrow 
\mathbb{R}^{+}$ over the multiplicative group $\mathbb{R}^{+}$ of scale
changes: 
\begin{equation}
\mathfrak{\hat{A}}(\mathcal{O})=\Gamma (\amalg _{\lambda \in \mathbb{R}^{+}}%
\mathfrak{A}_{\lambda }(\mathcal{O}))\ni \hat{A}:=(\mathbb{R}^{+}\ni \lambda
\longmapsto \hat{A}(\lambda )\in \mathfrak{A}_{\lambda }(\mathcal{O})).
\end{equation}%
\newline
The \textit{scaling algebra} $\hat{\mathfrak{A}}$ playing the role of the
global algebra is defined by the C*-inductive limit of all local algebras $%
\hat{\mathfrak{A}}(\mathcal{O})$ on the basis of the isotony $\hat{\mathfrak{%
A}}(\mathcal{O}_{1})\subset \hat{\mathfrak{A}}(\mathcal{O}_{2})$ for $%
\mathcal{O}_{1}\subset \mathcal{O}_{2}$.

The algebraic structures to make $\hat{\mathfrak{A}}(\mathcal{O})$ a unital
C*-algebra are defined in a \textit{pointwise manner} by 
\begin{eqnarray}
&&(c_{1}\hat{A}_{1}+c_{2}\hat{A}_{2})(\lambda ):=c_{1}\hat{A}_{1}(\lambda
)+c_{2}\hat{A}_{2}(\lambda ),  \notag \\
&&(\hat{A}_{1}\hat{A}_{2})(\lambda ):=\hat{A}_{1}(\lambda )\hat{A}%
_{2}(\lambda ),  \notag \\
&&\mathbf{\hat{1}}(\lambda ):=\mathbf{1}=\mathbf{1}_{\mathfrak{A}},  \notag
\\
&&(\hat{A}^{\ast })(\lambda ):=\hat{A}(\lambda )^{\ast },
\end{eqnarray}%
(for $\hat{A}_{1},\hat{A}_{2},\hat{A}\in \hat{\mathfrak{A}}(\mathcal{O})$, $%
c_{1},c_{2}\in \mathbb{C}$) and the C*-norm by%
\begin{equation}
||\,\hat{A}\,||:=\sup_{\lambda \in \mathbb{R}^{+}}\,||\,\hat{A}(\lambda
)\,||.
\end{equation}%
From the scaled actions $\mathfrak{A}_{\lambda }\underset{\alpha ^{(\lambda
)}}{\curvearrowleft }\mathcal{P}_{+}^{\uparrow }$ of the Poincar\'{e} group
on $\mathfrak{A}_{\lambda }$ with $\alpha _{a,\Lambda }^{(\lambda )}=\alpha
_{\lambda a,\Lambda }$, an action of $\mathcal{P}_{+}^{\uparrow }$ is
induced on $\mathfrak{\hat{A}}$ by 
\begin{equation}
(\hat{\alpha}_{a,\Lambda }(\hat{A}))(\lambda ):=\alpha _{\lambda a,\Lambda }(%
\hat{A}(\lambda ))\,,
\end{equation}%
\newline
in terms of which the essence of the condition (iii) is expressed simply as
the continuity of the action of the Poincar\'{e} group at its identity: $||\,%
\hat{\alpha}_{a,\Lambda }(\hat{A})-\hat{A}\,||\underset{(a,\Lambda
)\rightarrow (0,1)}{\rightarrow }0$. The so-defined scaling net\textit{\ }$%
\mathcal{O}\rightarrow \mathfrak{\hat{A}}(\mathcal{O})$ is shown to satisfy
all the properties to characterize a relativisitc local net of observables
if the original one $\mathcal{O}\rightarrow \mathfrak{A}(\mathcal{O})$ does.

Then, the scale transformation is defined by an automorphic action $\hat{%
\sigma}$ of the group $\mathbb{R}^{+}$ of scale changes on the scaling
algebra $\hat{\mathfrak{A}}$ given for $\forall \mu \in \mathbb{R}^{+}$ by 
\begin{equation}
(\hat{\sigma}_{\mu }(\hat{A}))(\lambda ):=\hat{A}(\mu \lambda ),\quad
\lambda >0\,,
\end{equation}%
satisfying the properties: 
\begin{eqnarray}
\hat{\sigma}_{\mu }(\hat{\mathfrak{A}}(\mathcal{O})) &=&\hat{\mathfrak{A}}%
(\mu \mathcal{O})\,,\quad \mathcal{O}\subset \mathbb{R}^{4}, \\
\hat{\sigma}_{\mu }\circ \hat{\alpha}_{a,\Lambda } &=&\hat{\alpha}_{\mu
a,\Lambda }\circ \hat{\sigma}_{\mu }\,,\quad (a,\Lambda )\in \mathcal{P}%
_{+}^{\uparrow }\,.
\end{eqnarray}

In this formulation, the roles of renormalization-group transformations to
relate observables at different scales are played by the scaling
transformations $\hat{\sigma}_{\mathbb{R}{^{+}}}$ acting isomorphically on
the scaling net\textit{\ }$\mathcal{O}\rightarrow \mathfrak{\hat{A}}(%
\mathcal{O})$. In view of the algebra $C(\mathbb{R}^{+})$ of scalar-valued
functions on $\mathbb{R}^{+}$ embedded in the centre of the scaling algebra $%
\mathfrak{\hat{A}}$, $C(\mathbb{R}^{+})\hookrightarrow \mathfrak{Z}(%
\mathfrak{\hat{A}})\subset \mathfrak{\hat{A}}$, it is no miracle for a 
\textit{broken} scale invariance caused by such \textit{explicit breaking}
terms as the mass $m$ to be restored as an \textquotedblleft
exact\textquotedblright\ symmetry described by an automorphic action $\hat{%
\sigma}$ of $\mathbb{R}^{+}$ when all the terms responsible for the explicit
breaking can be treated (as is common in practice) in terms of
scale-dependent classical variables like $\mathbb{R}^{+}\ni \lambda
\longmapsto m(\lambda )=\lambda ^{d_{m}}m_{0}$. It is also remarkable that
the final results obtained by Buchholz and Verch \cite{BucVer} through the
complicated analysis can naturally be seen just as a special case of the
previous definition of the augmented algebra $\mathfrak{\hat{F}}:=\Gamma
(G\times _{H}\mathfrak{F})$ for treating SSB with the choice of $H:=\mathcal{%
P}_{+}^{\uparrow }$, $G=H\rtimes \mathbb{R}^{+}$ (semidirect product of
groups) in combination with a slight modification due to the spacetime
dependence described by the local net structure: $\mathfrak{F}%
\Longrightarrow (\mathcal{O}\rightarrow \mathfrak{A}(\mathcal{O}))$ (upon
which the group $\mathbb{R}^{+}$ of scale changes acts). While this is the
case found in the vacuum situations which show the invariance under the
Poincar\'{e} group $\mathcal{P}_{+}^{\uparrow }=\mathbb{R}^{4}$ $\rtimes
L_{+}^{\uparrow }$ by definition, the typical thermal situations relevant to
our present contexts require more careful treatment because of the SSB of
Lorentz boost symmetry caused by temperatures. Here the Poincar\'{e} group $%
\mathcal{P}_{+}^{\uparrow }$ of relativisitc symmetry is broken down to $%
\mathbb{R}^{4}\rtimes SO(3)$ and, in the opposite direction, it is extended
to a larger one $\mathcal{P}_{+}^{\uparrow }\rtimes \mathbb{R}^{+}$
involving the \textit{broken} scale invariance, which may possibly be
extended at the critical points further to the conformal group $SO(2,4)$. It
is interesting to note that the series of group extensions $%
SO(3)\hookrightarrow SO(1,3)\hookrightarrow SO(1,3)\rtimes \mathbb{R}%
^{+}\hookrightarrow SO(2,4)$ (or its doube covering $SU(2)\hookrightarrow
SL(2,\mathbb{C})\hookrightarrow SL(2,\mathbb{C})\rtimes \mathbb{R}%
^{+}\hookrightarrow SU(2,2)$) can be understood in the context of the
Kantor-Koecher-Tits construction of Lie algebras associated with a symmetric
space and the corresponding Jordan triple system \cite{Sat80}, the last
member of which gives the analytic group of automorphisms of the tube domain 
$\mathbb{R}^{4}+iV_{+}\ni x^{\mu }+i\beta ^{\mu }$. If we start from the
choice of $H:=\mathcal{P}_{+}^{\uparrow }$, $G=H\rtimes \mathbb{R}^{+}$ even
in the thermal situation at $T\neq 0^{\circ }K$, the starting representation 
$(\pi ,\mathfrak{H})$ with $H$ as the group of unbroken symmetry should be
understood to contain already a non-trivial centre with $SO(3)\backslash
L_{+}^{\uparrow }\cong \mathbb{R}^{3}$ as its spectrum due to the SSB of $%
\mathcal{P}_{+}^{\uparrow }$ down to $\mathbb{R}^{4}\rtimes SO(3)$. In this
context, the scaled actions $\alpha _{a,\Lambda }^{(\lambda )}=\alpha
_{\lambda a,\Lambda }$ of Poincar\'{e} group on $\mathfrak{A}_{\lambda }$
can be naturally understood as the conjugacy change of the stability group $%
H\rightarrow gHg^{-1}$ from the point $He$ to $Hg^{-1}$ on the base space $%
H\backslash G=\mathbb{R}^{+}$: $s_{\lambda }(a,\Lambda )s_{\lambda
}^{-1}=(\lambda a,\Lambda )$, where $s_{\lambda }(x^{\mu })=\lambda x^{\mu }$%
.

\section{Scale changes on states}

In relation with the centre $\mathfrak{Z}(\mathfrak{\hat{A}})=\mathfrak{Z}(%
\mathfrak{\hat{A}}(\mathcal{O}))=C(\mathbb{R}^{+})$ arising from the broken
scale invariance, we have a canonical family of conditional expectations $%
\hat{\mu}$ from $\mathfrak{\hat{A}}$ to $\mathfrak{A}$ corresponding to
probability measures $\mu $ on $C(\mathbb{R}^{+})$: 
\begin{equation}
\hat{\mu}:\mathfrak{\hat{A}}\ni \hat{A}\longmapsto \int_{\mathbb{R}^{+}}d\mu
(\lambda )\hat{A}(\lambda )\in \mathfrak{A.}
\end{equation}%
(We can consider the case with $d\mu (\lambda )$ chosen to be the Haar
measure $d\lambda /\lambda $ of $\mathbb{R}^{+}$, which is, however, a
positive unbounded measure but not a probability one; the corresponding $%
\hat{\mu}$ becomes then an \textit{operator-valued weight }whose images are
not guaranteed to be finite.) By means of $\hat{\mu}$, any state $\omega \in
E_{\mathfrak{A}}$ can be lifted onto $\mathfrak{\hat{A}}$ by 
\begin{equation}
E_{\mathfrak{A}}\ni \omega \longmapsto \hat{\mu}^{\ast }(\omega )=\omega
\circ \hat{\mu}=\omega \otimes \mu \in E_{\mathfrak{\hat{A}}},
\end{equation}%
where we have used $\mathfrak{\hat{A}}\subset C(\mathbb{R}^{+},\mathfrak{A}%
)\cong \mathfrak{A}\otimes C(\mathbb{R}^{+})$.

In \cite{BucVer} the case of $\mu =\delta _{\lambda =1}$(: Dirac measure at
the identity of $\mathbb{R}^{+}$) is called a \textit{canonical lift }$\hat{%
\omega}:=\omega \circ \hat{\delta}_{1}$. The scale transformed state defined
by 
\begin{equation}
\hat{\omega}_{\lambda }:=\hat{\omega}\circ \hat{\sigma}_{\lambda }=\omega
\circ \hat{\delta}_{\lambda }
\end{equation}%
describes the situation at scale $\lambda $ due to the renormalization-group
transformation of scale change $\lambda $.

Conversely, starting from a state $\hat{\omega}$ of $\mathfrak{\hat{A}}$, we
can obtain its central decomposition as follows: first, we call two natural
embedding maps $\iota :\mathfrak{A\hookrightarrow \hat{A}}$ [$\left[ \iota
(A)\right] (\lambda )\equiv A$] and $\kappa :C(\mathbb{R}^{+})\simeq 
\mathfrak{Z}(\mathfrak{\hat{A}})\hookrightarrow \mathfrak{\hat{A}}$. Pulling
back $\hat{\omega}$ by $\kappa ^{\ast }:E_{\mathfrak{\hat{A}}}\rightarrow
E_{C(\mathbb{R}^{+})}$, we can define a probability measure $\rho _{\hat{%
\omega}}:=\kappa ^{\ast }(\hat{\omega})=\hat{\omega}\circ \kappa =$\ $\hat{%
\omega}\upharpoonright _{C(\mathbb{R}^{+})}$ on $\mathbb{R}^{+}$, namely, $%
\hat{\omega}\upharpoonright _{C(\mathbb{R}^{+})}(f)=\int_{\mathbb{R}%
^{+}}d\rho _{\hat{\omega}}(\lambda )f(\lambda )$ for $\forall f\in $ $C(%
\mathbb{R}^{+})$.

For any positive operator $\hat{A}=\int ad\hat{E}_{\hat{A}}(a)\in \mathfrak{%
\hat{A}}$, we can consider the central supports $c(\hat{E}_{\hat{A}}(\Delta
))\in Proj(\mathfrak{Z}_{\hat{\pi}_{\hat{\omega}}}(\mathfrak{\hat{A}}))$ of $%
\hat{E}_{\hat{A}}(\Delta )\in Proj(\hat{\pi}_{\hat{\omega}}(\mathfrak{\hat{A}%
})^{\prime \prime })$ with a Borel set $\Delta $ in $Sp(\hat{A})\subset
\lbrack 0,+\infty )$ satisfying $c(\hat{E}_{\hat{A}}(\Delta ))\hat{E}_{\hat{A%
}}(\Delta )=\hat{E}_{\hat{A}}(\Delta )$, from which we see that $\rho _{\hat{%
\omega}}^{\prime \prime }(c(\hat{E}_{\hat{A}}(\Delta )))=0$ implies $\hat{%
\omega}^{\prime \prime }(\hat{E}_{\hat{A}}(\Delta ))=0$, where $\hat{\omega}%
^{\prime \prime }$ and $\rho _{\hat{\omega}}^{\prime \prime }$ are the
extensions of $\hat{\omega}$ and $\rho _{\hat{\omega}}$ to $\hat{\pi}_{\hat{%
\omega}}(\mathfrak{\hat{A}})^{\prime \prime }$ and $L^{\infty }(\mathbb{R}%
^{+},d\rho _{\hat{\omega}})$, respectively. Thus, we can define the
Radon-Nikodym derivative $\omega _{\lambda }:=\frac{d\hat{\omega}}{d\rho _{%
\hat{\omega}}}(\lambda )$ of $\hat{\omega}$ w.r.t.~$\rho _{\hat{\omega}}$ as
a state on $\hat{\pi}_{\hat{\omega}}(\mathfrak{\hat{A}})^{\prime \prime }$
(in a similar way to \cite{Oza}) so that

\begin{equation}
\hat{\omega}(\hat{A})=\int d\rho _{\hat{\omega}}(\lambda )\omega _{\lambda }(%
\hat{A}(\lambda ))=\int d\rho _{\hat{\omega}}(\lambda )\omega _{\lambda }(%
\hat{\delta}_{\lambda }(\hat{A}))=\int d\rho _{\hat{\omega}}(\lambda )\left[
\omega _{\lambda }\otimes \hat{\delta}_{\lambda }\right] (\hat{A}).
\end{equation}%
Then, the pull-back $\iota ^{\ast }(\hat{\omega})=\hat{\omega}\circ \iota
\in E_{\mathfrak{A}}$ of $\hat{\omega}\in E_{\mathfrak{\hat{A}}}$ by $\iota
^{\ast }:E_{\mathfrak{\hat{A}}}\rightarrow E_{\mathfrak{A}}$ is given by 
\begin{equation}
\iota ^{\ast }(\hat{\omega})=\int d\rho _{\hat{\omega}}(\lambda )\omega
_{\lambda },
\end{equation}%
owing to the relation

\begin{equation}
\iota ^{\ast }(\hat{\omega})(A)=\hat{\omega}(\iota (A))=\int d\rho _{\hat{%
\omega}}(\lambda )\omega _{\lambda }(A)=\left[ \int d\rho _{\hat{\omega}%
}(\lambda )\omega _{\lambda }\right] (A).
\end{equation}%
Applying this relation to the scaled canonical lift, $\hat{\omega}_{\lambda
}:=\hat{\omega}\circ \hat{\sigma}_{\lambda }=(\omega \circ \hat{\delta}%
_{1})\circ \hat{\sigma}_{\lambda }=\omega \circ \hat{\delta}_{\lambda }$, of
a state $\omega \in E_{\mathfrak{A}}$, we can easily see $\iota ^{\ast
}(\omega \circ \hat{\delta}_{\lambda })=\iota ^{\ast }(\hat{\omega}_{\lambda
})=\omega _{\lambda }[=\frac{d\hat{\omega}_{\lambda }}{d\delta _{\lambda }}%
(\lambda )]=\phi _{\lambda }(\omega )$, where $\phi _{\lambda }$ is the
isomorphism introduced in \cite{BucVer} between $\omega $ and the canonical
lift $\hat{\omega}_{\lambda }\in E_{\mathfrak{\hat{A}}}$ projected onto $%
\mathfrak{\hat{A}}/\mathrm{ker}(\hat{\pi}_{\hat{\omega}}\circ \hat{\sigma}%
_{\lambda })$.

Thus, we can lift canonically any state $\omega \in E_{\mathfrak{A}}$ from $%
\mathfrak{A}$ to $\hat{\omega}\in E_{\mathfrak{\hat{A}}}$, and, after the
scale shfit $\hat{\sigma}_{\lambda }$ on $\mathfrak{\hat{A}}$, return $\hat{%
\omega}\circ \hat{\sigma}_{\lambda }$ back onto $\mathfrak{A}$: $\phi
_{\lambda }(\omega )=\omega _{\lambda }=\iota ^{\ast }(\omega \circ \hat{%
\delta}_{\lambda })$, as result of which we obtain the scaled-shifted state $%
\omega _{\lambda }\in E_{\mathfrak{A}}$ from $\omega \in E_{\mathfrak{A}}$ 
\textit{in spite of the absence of scale invariance on} $\mathfrak{A}$.

Now applying this procedure to $\omega =\omega _{\beta }$ (: any state
belonging to the family of relativistic KMS states with the same $(\beta
^{2})^{1/2}$), we have a genuine KMS state by going to their rest frames.
Then we have $\hat{\omega}_{\lambda }=(\widehat{\omega _{\beta }})_{\lambda
}=\omega _{\beta }\circ \widehat{\delta _{\lambda }}$ which is shown to be a
KMS state at $\beta /\lambda $:%
\begin{eqnarray}
(\omega _{\beta }\circ \widehat{\delta _{\lambda }})(\hat{A}\hat{\alpha}_{t}(%
\hat{B})) &=&\omega _{\beta }(\hat{A}(\lambda )\alpha _{\lambda t}(\hat{B}%
(\lambda )))  \notag \\
=\omega _{\beta }(\alpha _{\lambda t-i\beta }(\hat{B}(\lambda ))\hat{A}%
(\lambda )) &=&\omega _{\beta }(\alpha _{\lambda (t-i\beta /\lambda )}(\hat{B%
}(\lambda ))\hat{A}(\lambda ))  \notag \\
&=&(\omega _{\beta }\circ \widehat{\delta _{\lambda }})(\hat{\alpha}%
_{t-i\beta /\lambda }(\hat{B})\hat{A}),
\end{eqnarray}%
and hence, $(\widehat{\omega _{\beta }})_{\lambda }\in \hat{K}_{\beta
/\lambda }$, $\phi _{\lambda }(\omega _{\beta })\in K_{\beta /\lambda }$.

As has been already remarked, the above discussion is seen to apply equally
to the spontaneous as well as \textit{explicitly broken} scale invariance
with such \textit{explicit breaking} parameters as mass terms. The actions
of scale transformations on such variables as $x^{\mu }$, $\beta ^{\mu }$
and also conserved charges are just straightforward, which is justified by
such facts that the first and the second ones are of \textit{kinematical}
nature and that the second and the third ones exhibit themselves in the 
\textit{state labels} for specifying the relevant \textit{sectors} in the
context of the superselection structures \cite{BOR01, Unif03}. This gives an
alternative verification to the so-called \textit{non-renormalization
theorem of conserved charges}. In sharp contrast, other such variables as
coupling constants (to be read off from the data of correlation functions or
Green's functions) are affected by the \textit{scaled dynamics,} and hence,
may show non-trivial scaling behaviours with deviations from the canonical
(or kinematical) dimensions, in such forms as the running couplings or
anomalous dimensions. Thus, the transformations $\hat{\sigma}_{\lambda }$
(as \textquotedblleft exact\textquotedblright\ symmetry on the augmented
algebra $\mathfrak{\hat{A}}$) are understood to play the roles of the
renormalization-group transformations (as broken symmetry on the original
algebra $\mathfrak{A}$).

As a result, we see that \textit{classical macroscopic observable} $\beta $
naturally emerging from a microscopic quantum system is verified to be an 
\textit{order parameter }of\textit{\ broken scale invariance} involved in
the renormalization group in relativisitc QFT.

\section{Summary and outlook}

To equip such expressions as \textquotedblleft broken scale
invariance\textquotedblright\ and its \textquotedblleft order
parameter\textquotedblright\ with their precise formulations, we have
adopted a scheme to incorporate \textit{spontaneously as well as explicitly
broken} symmetries with the criterion for symmetry breakdown on the basis of
an augmented algebra with a non-trivial centre in such forms as $\mathfrak{%
\hat{F}}=\Gamma (G\times _{H}\mathfrak{F)}$ or $\mathcal{O}\longmapsto 
\mathfrak{\hat{A}}(\mathcal{O})=\Gamma (\amalg _{\lambda \in \mathbb{R}^{+}}%
\mathfrak{A}(\lambda \mathcal{O}))$, the latter of which is just a
re-interpretation of the Buchholz-Verch scaling net of local observables
adapted to the former. As an algebra of the \textit{composite system} of a
genuine quantum one together with classical macroscopic one (embedded as the
centre), the augmented algebra $\mathfrak{\hat{F}}$ or $\mathfrak{\hat{A}}$
can play such important roles that

\begin{itemize}
\item[a)] it allows a symmetry \textit{broken explicitly} by breaking terms
(such as non-vanishing masses in the case of scale invariance) to be
formulated in terms of the \textit{symmetry transformations} acting on $%
\mathfrak{\hat{A}}$ through automorphisms $\in Aut(\mathfrak{\hat{A}})$
which is realized by the simultaneous changes of the breaking terms
belonging to the centre $\mathfrak{Z}(\mathfrak{\hat{A}})$ to cancel the
breaking effects,

\item[a')] for a \textit{spontaneously broken} symmetry, this augmented
algebra naturally accommodates its \textit{covariant} unitary representation
as an induced representation from a subgroup of the remaining unbroken
symmetry (at the expense of the non-trivial centre characterizing the
symmetry breaking),

\item[b)] the \textit{continuous} behaviours of order parameters under the
broken symmetry transformations is algebraically expressed at the C*-level
of the (C*-algebraic) centre $\mathfrak{Z}(\mathfrak{\hat{A}})$ in sharp
contrast to the \textit{discontinuous} ones at the W*-level $\mathfrak{Z}%
_{\pi }(\mathfrak{\hat{A}})$ of representations owing to the mutual
disjointness among representations corresponding to different values of
order parameters (as points on $Spec(\mathfrak{Z}_{\pi }(\mathfrak{A}))$).
To this continuous order parameter some \textit{external field} can further
be coupled, like the coupling between the magnetization and an external
magnetic field in the discussion of a Heisenberg ferromagnet. Using this
coupling, we can examine, for instance, the mutual relations between the
magnetization caused by an external field and the spontaneous one, the
latter of which persists in the asymptotic removal of the former in
combination with the hysteresis effects. Without introducing the augmented
algebra $\mathfrak{\hat{A}}$, it seems difficult for this kind of
discussions to be adapted to the case of QFT.
\end{itemize}

Then, the mutual relation between states on $\mathfrak{A}$ and $\mathfrak{%
\hat{A}}$ is clarified, on the basis of which the verification of the
statement on the behaviour of the inverse temperatures is just reduced to a
simple computation of checking the parameter shift occurring in the KMS
condition under the scale transformation. What is interesting in this
observation about the roles of the (inverse) temperature $\beta $ is that it
exhibits the \textit{cross over between thermal and geometric aspects}
expressed in $\beta ^{\mu }=\beta u^{\mu }$ and in the spacetime
transformations $\mathcal{P}_{+}^{\uparrow }\rtimes \mathbb{R}^{+}$
including the scale one, respectively.

While the above scale transformations in the \textit{real} version cause
actual changes on equilibrium states with a temperature into another, there
is also a \textit{virtual} version which does \textit{not} change the real
temperature but which acts in the virtual direction of the \textit{%
interpolating family of non-commutative }$L^{p}$\textit{-spaces} \cite%
{AraMas} to be built here on the von Neumann algebras $\mathfrak{M}:=\pi
_{\beta }(\mathfrak{A})^{\prime \prime }$, defined as the GNS representation
associated to a KMS state $\omega _{\beta }$. A \textit{virtual temperature }%
$\tau $\textit{\textbf{\ }}associated with an $L^{p}$\textit{-}space\textit{%
\ }$L^{p}(\mathfrak{M};\omega _{\beta })$ is given here as $0\leq \tau
=\beta /p\leq \beta $ (or, $T=(k_{B}\beta )^{-1}\leq pT\leq \infty $: i.e.,
high temperature side) because of the restriction of $1\leq p\leq \infty $.
From the physical viewpoint, this restriction can be naturally understood as
the difficulty in moving kinematically to the direction of lower
temperatures caused by the possible occurrence of \textit{phase transitions}
which cannot be treated without the dynamical considerations.

This context of interpolation theory involves various interesting aspects,
such as the extension of some key notions in the (quantum) information
geometry and statistical inference theory like $\alpha $-divergences,
relative entropy, Fisher information, etc., which have been traditionally
treated in quantum systems with finite degrees of freedom, or even in the
finite-dimensional matrix algebras \cite{Ama85, Has93}. On the present
setting-up, it is quite natural to extend the discussions in this context to
a general quantum dynamical system with infinite degrees of freedom. For
instance, with $\alpha :=\frac{1}{q}-\frac{1}{p}\neq \pm 1$, $\frac{1}{p}+%
\frac{1}{q}=1$, we have an $\alpha $-divergence%
\begin{equation}
D^{(1/q-1/p)}(T_{1}||T_{2})=pq\left[ \frac{\left\vert \left\vert
T_{1}\right\vert \right\vert _{p}^{p}}{p}+\frac{\left\vert \left\vert
T_{2}\right\vert \right\vert _{q}^{q}}{q}-[T_{1},T_{2}]_{\phi _{0}}\right] ,
\end{equation}%
where $[T_{1},T_{2}]_{\phi _{0}}$ is the pairing between $L^{p}$ and $L^{q}$
w.r.t.~a faithful normal semifinite weight $\phi _{0}$ (or, a faithful
normal state such as $\omega _{\beta }$) for $T_{1}=u_{1}\Delta _{\varphi
_{1},\phi _{0}}^{1/p}\in L^{p}(\mathfrak{M},\phi _{0})$, $T_{2}=u_{2}\Delta
_{\varphi _{2},\phi _{0}}^{1/q}\in L^{q}(\mathfrak{M},\phi _{0})$, with $%
u_{i}$: partial isometries and $\Delta _{\varphi _{i},\phi _{0}}$: relative
modular operator from $\phi _{0}$ to $\varphi _{i}\in \mathfrak{M}_{\ast ,+}$%
. The problems related with this geometric aspect of temperature states will
be discussed elsewhere.

\section*{Acknowledgement}

I would like to express my sincere thanks to Prof.~Takahiro Kawai for his
encouragements in my reseach programmes.

\end{document}